\documentclass[letterpaper,twocolumn,10pt]{article}
\usepackage{usenix,epsfig,endnotes}

\usepackage{cite}
\usepackage[english]{babel}
\usepackage{graphicx}
\usepackage[T1]{fontenc}
\usepackage[latin1]{inputenc}
\usepackage[psnames,svgnames]{xcolor}
\usepackage[justification=centering, caption=false]{subfig}
\usepackage{amsmath,amssymb,amsthm,latexsym}
\usepackage{algorithm}
\usepackage{algpseudocode} 
\usepackage{xspace}
\usepackage{cite}
\usepackage{enumitem}
\usepackage[draft]{hyperref}

\hypersetup{%
pdftitle={Fast Product-Matrix Regenerating Codes},
pdfauthor={Nicolas Le Scouarnec},
pdfkeywords={practical, regenerating codes,  system, implementation, product-matrix},
pdfstartview={FitH},
urlcolor=darkblue,
pdfborder=0 0 0,
bookmarksnumbered
}%

\begin{document}

%% Target https://www.usenix.org/conference/fast15/call-for-papers

%don't want date printed

\title{Fast Product-Matrix Regenerating Codes}

\author{
%{\rm Anonymous}
	{\rm Nicolas Le\ Scouarnec}\\
	Technicolor, France
%	{\rm 5 September 2014}
	%\IEEEauthorblockN{Nicolas Le Scouarnec}
	%\IEEEauthorblockA{Technicolor, Rennes, France}
}

\maketitle

% Use the following at camera-ready time to suppress page numbers.
% Comment it out when you first submit the paper for review.
%\thispagestyle{empty}

\subsection*{Abstract}
Distributed storage systems support failures of individual devices by the use of replication or erasure correcting codes. While erasure correcting codes offer a better storage efficiency than replication for similar fault tolerance, they incur higher CPU consumption, higher network consumption and higher disk I/Os.  To address these issues, codes specific to storage systems have been designed. Their main feature is the ability to repair a single lost disk efficiently. In this paper, we focus on one such class of codes that minimize network consumption during repair, namely regenerating codes. We implement the original Product-Matrix Regenerating codes as well as a new optimization we propose and show that the resulting optimized codes allow achieving 790 MB/s for encoding in typical settings. Reported speeds are significantly higher than previous studies, highlighting that regenerating codes can be used with little CPU penalty.
%\end{abstract}

\section{Introduction}
The increasing needs for storage in datacenter pushed by numerous cloud-based storage services has lead to the development of storage systems with a better tradeoff between reliability and storage. The usual solution  is to rely on erasure correcting codes~\cite{Dimakis2010a,Fikes2010,Ford2010,Calder2011,Rashmi2013}. While they allow decreasing storage costs, they generally come with higher costs in term of network, I/O or CPU and thus hardware cost and power consumption. To alleviate these drawbacks, optimized codes have been designed: \emph{(i)} regenerating codes minimize network-related costs~\cite{Dimakis2010, Dimakis2010a,Rashmi2011}, \emph{(ii)} locally repairable codes minimize  I/O related costs~\cite{Gopalan2011,Huang2007, Papailiopoulos2012a, Khan2012}, \emph{(iii)} other codes minimize CPU related costs~\cite{Plank2011, Plank2012, Luo2013}. In the rest of the paper, we will focus on regenerating codes as they offer the best tradeoff between network and storage. 

Regenerating codes have been mainly studied with respect to either theoretical aspects (\emph{i.e.}, existence)~\cite{Dimakis2010, Dimakis2010a,Rashmi2011,Cadambe2011a,Shah2010b,Shah2012,Suh2011,Suh2010a,Tamo2011,Tamo2012}; or cost in bandwidth~\cite{Li2013,Hu2014,Hu2011}. Beside these studies, few have looked at system aspects, including encoding/decoding throughput (i.e., CPU cost). The throughputs reported are 50 KB/s (k=32, d=63) in~\cite{Duminuco2009}, 0.6 MB/s (k=16, d=30) in~\cite{Jiekak2013}, 100MB/s (k=2, d=3)~\cite{Hu2014}. So far, reported speeds, except for very low values of $k$, are incompatible with practical deployments. 
%One of these previous study~\cite{Jiekak2013} reported that Product Matrix compared fairly to Reed-Solomon codes, and where much faster than random linear regenerating codes (such as the one used in~\cite{Duminuco2009} and~\cite{Hu2014}) but without reporting achievable throughput as they relied on a Java-based implementation which lacked optimized support for Galois Field and bit-level operations. 

In this paper, we go beyond these, and provide insights useful to practical deployment of regenerating codes.
\begin{itemize}
\item{} We describe an optimization that almost quadruple the performance of product-matrix codes~\cite{Rashmi2011} (\emph{e.g.}, from 210 to 790 MB/s for $k=8$ systematic codes) % 
%and lower impact of repair on disks. (by 40\% for k=8).  
\item{} We report throughputs for product-matrix regenerating codes and compare them to Reed-Solomon codes. For systematic codes with $k=8$, our optimized product-matrix codes encode at 790 MB/s when Reed-Solomon codes encode at 1640 MB/s.
%\item{} We describe the single step repair procedure we use for Reed-Solomon codes so as to save CPU cost (decoding and then encoding processed data at XX MB/s while the single step repair can be performed at XX MB/s). This allows us to compare Reed-Solomon and Regenerating codes on a fair basis with respect to CPU-related repair costs.
\end{itemize}

Section~\ref{sec:back} describes some background on product-matrix regenerating codes as well as libraries we rely on for our implementation (\emph{i.e.}, Jerasure~\cite{Plank2014} and GF-Complete~\cite{Plank2013}). Section~\ref{sec:main} describes the transformations we apply to product-matrix codes to turn them into linear codes, or to enhance the performance of the systematic form by sparsifying the encoding matrix. Section~\ref{sec:eval} shows the performance achieved and studies the impact of the various parameters.

\section{Background}
\label{sec:back}
Fault tolerance mechanisms such as replication or erasure correcting codes are used to limit the impact of failures in distributed storage systems. An erasure correcting code encodes $k$ blocks of original data (column vector $X$) to $n$ blocks of encoded data (column vector $Y$), so that the $k$ original blocks can be recovered from any $k$ encoded blocks. The code can be defined by its generator matrix G of dimension $n\times{}k$. The encoding operation is then expressed as $Y=GX$; and the decoding operation as $X=\tilde{G}^{-1}\tilde{Y}$ where $\tilde{G}$ (resp. $\tilde{Y}$) are the rows of $G$ (resp. $Y$) that correspond to the remaining encoded blocks. Each device stores one block of data.

A key feature of distributed storage systems is their ability to self-repair (\emph{i.e.}, to recreate lost encoded blocks whenever some device fails). Erasure correcting codes support such repair by decoding $k$ blocks and encoding again. This implies the transfer of $k$ blocks to recreate a single lost block. leading to high network-related repair cost. To solve this issue, regenerating codes~\cite{Dimakis2010} have been designed as randomized codes offering an optimal tradeoff between storage and network bandwidth (\emph{i.e.}, network-related repair cost). Then, numerous studies~\cite{Dimakis2010a,Cadambe2011a,Shah2010b,Shah2012,Suh2011,Suh2010a,Tamo2011,Tamo2012} have focused on deterministic regenerating codes including product-matrix codes which are applicable to a wide set of parameters~\cite{Rashmi2011}. 

\begin{figure}[t]
\centering
\includegraphics[width=0.6\linewidth]{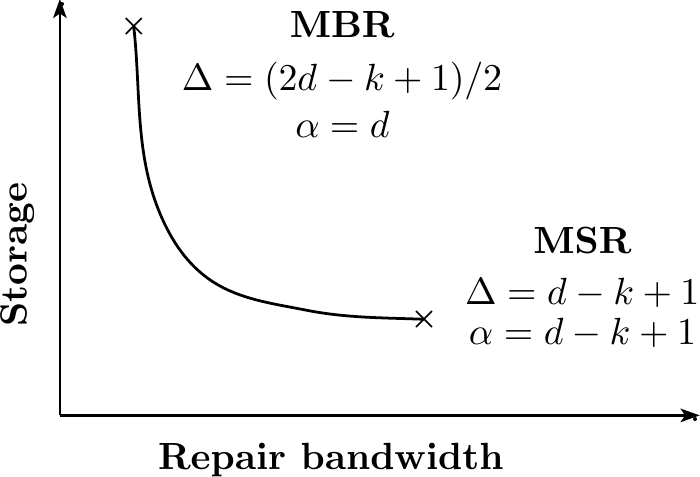}
\caption{The tradeoff curve of regenerating codes with Minimum Storage Regenerating codes (MSR) and Minimum Bandwidth Regenerating codes (MBR)}
\label{fig:msrmbr}
\end{figure}

Regenerating codes achieve the optimal tradeoff between storage and bandwidth with two main variants: \emph{(i)} MSR codes that favor storage, and \emph{(ii)} MBR codes that favor bandwidth, as depicted on Figure~\ref{fig:msrmbr}. MSR codes encode $k\Delta$ blocks to $n\alpha$ blocks with $\alpha=\Delta=d-k+1$. Each device stores $\alpha$ blocks. During repair, $d$ blocks are transferred over the network. MBR are similar but have $\alpha = d$ and $\Delta = (2d-k+1)/2$ thus resulting in higher storage costs but lower network bandwidth usage.

Regenerating codes can be implemented using linear codes. Similarly to regular erasure correcting codes, the code can be defined by its generator matrix $G$ of dimension $n\alpha \times k\Delta$. The encoding operation is then expressed as $Y=GX$ (where $X$ is a column vector of $k\Delta$ blocks and $Y$ is a column vector of $n\alpha$ blocks); and the decoding operation as $X=\tilde{G}^{-1}\tilde{Y}$ where $\tilde{G}$ (resp. $\tilde{Y}$) are the rows of $G$ (resp. $\tilde{Y}$) that correspond to the remaining encoded blocks. The factor $\alpha$ in dimensions comes from the fact that each device stores $\alpha$ instead of $1$ block for erasure correcting codes\footnote{This allows devices using regenerating codes to repair $\alpha$ blocks by downloading $1$ block from $d$ devices thus saving bandwidth when compared to erasure correcting codes that repair $1$ block by downloading $1$ block from $k$ devices.}. As a consequence, the encoding and decoding complexities come with an additional factor $\alpha\Delta$. Given that in a typical setting $\alpha \propto k$ and $\Delta \propto k$, the encoding and decoding complexities, which are $\Omega(k^2)$ for erasure correcting codes, become $\Omega(k^4)$ for regenerating codes.

Product-matrix regenerating codes~\cite{Rashmi2011} rely on specific encoding and decoding algorithms and are not defined using a generator matrix as usual linear codes. The $k\Delta$ original blocks of $X$ are arranged (with repetitions and zeros) into a message matrix $M$ of dimension $d \times \alpha$ in a way specific to MSR or MBR. The code is defined by its encoding matrix $\Psi$ of dimension $n \times d$. The encoding operation is thus defined as $C = \Psi{}M$ such that $C$ is a $n \times \alpha$ matrix containing the $n\alpha$ encoded blocks. The decoding algorithms are specific to the type of regenerating code used (MSR or MBR) and are described in~\cite{Rashmi2011}.

% For product-matrix codes, this comes down to applying the decoding operation on $X$ arranged as a $k \times \alpha$ matrix to obtain a precoded message matrix $M'$ and encoding it as $C = \Psi{}M'$. The rest of the paper will focus on the performance of systematic product-matrix codes.

Regenerating codes have been mainly studied for their saving in bandwidth, leaving aside computational performance. A few papers have implemented and measured the performance of various regenerating codes such as randomized codes in~\cite{Duminuco2009,Jiekak2013,Hu2014}. They achieve around 50 KB/s using a pure Java-based implementation (k=32, d=63) in~\cite{Duminuco2009,Jiekak2013}, 100MB/s for very small codes ($k=2$, $d=3$) in~\cite{Hu2014}. Implementation of deterministic codes achieves relatively higher speed such as 0.6 MB/s (k=16, d=30) in pure Java~\cite{Jiekak2013}. 
%Finally, FIGURE OUT WHAT CORE IS ( http://arxiv.org/pdf/1302.3344v2.pdf ) AND WHAT PERFORMANCE RESULTS CAN BE EXTRACTED FROM THEIR BENCHMARKS (USE MULTITHREADED CODING  ON A QUAD CORE, 3.1 Ghz we use SINGLE CORE, 2.5 Ghz) NOT CLEAR ON RECONSTRUCTION / REPAIR / RECOVERY.... 
Encoding throughput as reported in these publications is a factor that is likely to limit the deployment of regenerating codes, since saving network bandwidth increase a lot processing costs. 

Regarding the implementation of erasure correcting codes, highly optimized libraries have been recently released and provide fast arithmetic on Galois Field, namely GF-Complete~\cite{Plank2013} or linear codes, namely Jerasure 2.0~\cite{Plank2014}. We will use these two libraries in our implementation of product-matrix regenerating codes.

\section{Fast PM Regenerating codes}
\label{sec:main}

We now describe the key aspects of our implementation of product matrix regenerating codes. Basically, the best performance is achieved by turning the PM codes into systematic linear codes, and using the sparse code we define in this paper instead of the \emph{vanilla} code from the seminal paper~\cite{Rashmi2011}. The performance impacts of choosing each optimization are detailed in Section~\ref{sec:eval}.

\subsection{Linearization of Product-Matrix codes}
\label{ssec:linear}
When implementing product matrix regenerating codes, two alternatives are possible, namely \emph{(i)} applying the algorithms described in the seminal paper~\cite{Rashmi2011} or \emph{(ii)} transforming them to linear codes and using generic algorithms for linear codes such as the ones implemented in Jerasure~\cite{Plank2014}. To support our evaluation of both strategies in Section~\ref{sec:eval}, we explain how product matrix codes can be transformed into equivalent linear codes.

To obtain the linearized version of the product matrix code, we need to create a $n\alpha \times k\Delta$ generator matrix $G$ for a given code. First, we construct an index matrix $L$ accordingly to the definition of the message matrix $M$ of the seminal paper, so that $M_{i,j} = X_{L_{i,j}}$. For example, for an MSR code with $k=3, \alpha=\Delta=2$, we would have the index matrix $L$ of dimension ${2\alpha} \times {\alpha}$ such that
\begin{equation*}
L = \left[ 
\begin{matrix}
  1 & 2  & 4 & 5\\
  2 & 3  & 5 & 6\\
 \end{matrix}
\right]^\mathrm{t}
\end{equation*}

Encoding by the product-matrix regenerating code is defined as $C = \Psi M$. The element $C_{i,j}$ and the corresponding element $Y_{\alpha{}i+j}$ for linear regenerating code defined as  $Y = GX$  are computed using %
\begin{align*}
	C_{i,j} = \sum_{l=1}^n \Psi{}_{i,l} X_{L_{l,j}} && 	Y_{\alpha{}i+j} = \sum_{l'=1}^{k\delta} G_{\alpha{}i+j,l'} X_{l'}
\end{align*}

For all $i, j$, we have $C_{i,j} = Y_{\alpha{}i+j}$. Applying a change of variable $l' = L_{l,j}$, and noticing that \emph{(i)} by construction (see~\cite{Rashmi2011}) of $M$ and $L$, no row nor column of $L$ has duplicates values, \emph{(ii)} the equality can hold only if $G_{\alpha{}i+j,l'} = 0$ for any $X_{l'}$ not present on the left-hand side, we obtain
\begin{align*}
\sum_{l=1}^n \Psi{}_{i,l} X_{L_{l,j}} =  \sum_{l=1}^{n} G_{\alpha{}i+j,L_{l,j}} X_{L_{l,j}}
\end{align*}

This implies that the generator matrix $G$ such that
\begin{align*}
G_{\alpha{}i+j,l'} = \left\{
\begin{matrix}
 \Psi{}_{i,l} & \textrm{if there exist $l$ such that $l' = L_{l,j}$}\\
 0               & \textrm{otherwise}
\end{matrix}
\right.
\end{align*}
is equivalent to a product matrix code (MSR or MBR) defined by $\Psi$ and $L$.

We explore the benefits of linearizing product matrix codes in Section~\ref{sec:eval}, and see that it leads to significant improvement when using systematic codes. Also, as side advantages, linearization allows \emph{(i)} re-using libraries designed for regular erasure codes thus significantly, \emph{(ii)} simpler \emph{zero-copy/lazy} (\emph{i.e.}, decoding only lost blocks) implementation as it is a single step transformation.

\subsection{Systematic codes}
A key feature of codes for storage, being erasure correcting codes or regenerating codes, is their ability to be transformed into systematic codes. A systematic erasure correcting code (resp. systematic regenerating code) is a code such that the $k$ (resp. $k\Delta$) first blocks of $Y$ are equal to the original data $X$.
Systematic codes have two main advantages: \emph{(i)} accessing data does not require decoding provided that none of the $k$ first devices has failed thus enhancing the performance of the common case\footnote{Even if the system is designed to tolerate failures, most of the time, the device storing the data accessed is likely to be available.}, \emph{(ii)} encoding tends to be less costly as we only need to compute the $n-k$ (resp. $n\alpha-k\Delta$) last rows of $Y$, the $k$ (resp. $k\Delta$) first being equal to $X$ by construction. 

The construction of systematic codes is based on the fact that encoding and decoding operations are more or less commutative. Hence, one can apply the decoding to the original data $X$ to obtain precoded data $Z$ and then encoding $Z$ to obtain $Y$ such that the $Y_{[1\dots{}k\alpha]} = X$.
For linear codes, the encoding is thus defined as $Y = G\tilde{G}^{-1}X$ with $\tilde{G}$ being the $k\Delta$ first rows of $G$. The resulting systematic code has a generator matrix $G' = G\tilde{G}^{-1}$, such that $G' = \left[\begin{matrix}I & {G''}^t\end{matrix}\right]^t$. By construction, $G'$ inherits appropriate properties (\emph{e.g.}, all $k\times{}k$ (resp. $k\alpha\times{}k\alpha$) sub-matrices are full-rank) from $G$ ensuring that it is an appropriate generator matrix. An alternative could be to choose $G''$ such that $G' = \left[\begin{matrix}I & {G''}^t\end{matrix}\right]^t$ has the needed properties. 

While building $G'$ directly is possible for some codes, this is not easy for MSR product-matrix regenerating codes since matrix $\Psi$ (from which $G$ can be derived as explained in Section~\ref{ssec:linear}) must satisfy several constraints  to allow efficient repair. Hence, the seminal paper builds systematic MSR codes by successively decoding and encoding.  For MBR product-matrix codes, since the matrix $\Psi$ is less constrained, the seminal paper~\cite{Rashmi2011} directly gives a systematic encoding matrix $\Psi = \left[\begin{matrix} I & {\Psi''}^t\end{matrix}\right]^t$ thus leading to simpler efficient implementation.

However, as we explore in Sub-section~\ref{ssec:sparse} and Section~\ref{sec:eval}, such indirect construction of Product-Matrix MSR codes has a significant impact on the computing cost thus requiring optimization beyond the construction from paper~\cite{Rashmi2011} as presented in the next sub-section.

\subsection{Sparse MSR PM codes}
\label{ssec:sparse}
The paper~\cite{Rashmi2011} specifies the following constraints on the encoding matrix $\Psi$ for MSR product-matrix codes.
\begin{itemize}[noitemsep,nolistsep]
\item $\Psi = \left[ \Phi  \;\; \Lambda\Phi \right]$ where $\Phi$ is a $n\times{}\alpha$ matrix and $\Lambda$ is a $n\times{}n$ diagonal matrix. %
\item Any $d$ rows of $\Psi$ are linearly independent. %
\item Any $\alpha$ rows of $\Phi$ are linearly independent. %
\item All values of $\Lambda$ are distinct. %
\end{itemize} 

The construction suggested in the paper is to take $\Psi$ as a Vandermonde matrix (\emph{i.e.}, $\Phi_{i,j} = g^{(i-1)(j-1)}$) that satisfies all needed properties. For example, if $n=5, k=3, d=4, \alpha=2$ and the finite field used has a generator element $g$, this gives the following matrices:\\[-5eX]

\def\nesp{\!\!\!\!\!\!\!\!\!\!\!\!}
\def\nespb{\!\!\!\!}
\footnotesize
\begin{align*}
\Phi\!=\!\begin{pmatrix}
1\nespb  & 1  \\
1\nespb & g^1  \\
1\nespb & g^2  \\
1\nespb & g^3  \\
1\nespb & g^4 
\end{pmatrix}
&&
\Lambda\!=\!\begin{pmatrix}
\!\!\!1\nesp&\!\!\!  &\!\!\!  &\!\!\!  &\!\!\! &\!\!\! \\
\!\!\! & g^2\nesp &\!\!\!  \!\!\!&\!\!\! &\!\!\! &\!\!\! \\
\!\!\! &\!\!\!  & g^4\nesp &\!\!\! &\!\!\! &\!\!\!  \\
\!\!\! &\!\!\!  &\!\!\!  & g^6\nesp &\!\!\! & \!\!\!\\
\!\!\! &\!\!\!  & \!\!\! & \!\!\! & g^8\nesp  & \!\!
\end{pmatrix}
&&
\Psi\!=\!\begin{pmatrix}
1\nespb & 1\nespb & 1\nespb &  1\nespb \\
1\nespb & g^1\nespb & g^2\nespb & g^3\nespb \\
1\nespb & g^2\nespb & g^4\nespb & g^6\nespb \\
1\nespb & g^3\nespb & g^6\nespb & g^9\nespb \\
1\nespb & g^4\nespb & g^8\nespb & g^{12}
\end{pmatrix}
\end{align*}
\normalsize

Interestingly, such construction based on dense encoding matrix $\Psi$ leads to sparse generator matrix $G$ as shown on Table~\ref{tab:sparse} (e.g., $G$ contains 75\% of zeros for $k=8$). Sparsity is important as multiplications by zero can be skipped resulting in lower computational costs.  However, when the code is turned into systematic form, sparsity is lost and the lower part of the resulting generator $G''$ contains 0~\% of zero.
% This explains that product-matrix regenerating codes perform rather well when compared to Reed-Solomon codes even if their generator matrix are bigger by a factor $\alpha$ as reported in~\cite{Jiekak2013}.

In order to reduce the computational cost of product-matrix MSR codes, we propose an alternative construction which satisfies the conditions aforementioned but gives slightly sparser $\Psi$ and $G$, and much sparser systematic generator $G''$. We take $\Phi$ as an identity matrix concatenated to a Cauchy-Matrix (\emph{i.e.}, $\Phi_{i,j} = (g^{i+\alpha} - g^{j-1})^{-1}$ for all $i > \alpha$. $\Lambda$ is set to $\Lambda_{i,i} = \frac{g^{i+\alpha}-g^{0}}{g^{i+\alpha}-g^{\alpha}}$ and $\Lambda_{i,j} = 0$ if $i \neq j$. For $n=5, k=3, d=4, \alpha=2$ and a finite field having a generator element $g$, this gives the following matrices:\\[-5eX]

\def\esp{\;\;\;\;\;\;}
\footnotesize
\begin{align*}
\Phi\!=\!\begin{pmatrix}
1\nespb  & 0  \\
0\nespb & 1  \\
\frac{1}{g^{2\alpha+1}-g^0}\nespb &  \frac{1}{g^{2\alpha+1}-g^1} \\
\frac{1}{g^{2\alpha+2}-g^0}\nespb &  \frac{1}{g^{2\alpha+2}-g^1} \\
\frac{1}{g^{2\alpha+3}-g^0}\nespb &  \frac{1}{g^{2\alpha+3}-g^1} 
\end{pmatrix}
&&
\Lambda\!=\!\begin{pmatrix}
\frac{g^{\alpha+1}-g^0}{g^{\alpha+1}-g^{\alpha}}\esp\esp\esp\esp\\\
\esp\frac{g^{\alpha+2}-g^0}{g^{\alpha+2}-g^{\alpha}}\esp\esp\esp\\
\esp\esp\frac{g^{2\alpha+1}-g^0}{g^{2\alpha+1}-g^{\alpha}}\esp\esp \\
\esp\esp\esp\frac{g^{2\alpha+2}-g^0}{g^{2\alpha+2}-g^{\alpha}}\esp\\
\esp\esp\esp\esp\frac{g^{2\alpha+3}-g^0}{g^{2\alpha+3}-g^{\alpha}} 
\end{pmatrix}
\end{align*} %
\vspace{-0.5cm}\begin{align*}
\Psi\!=\! \left[\Phi \esp  \Lambda\Phi\right]
%\begin{pmatrix}
%1\nespb & 0\nespb & \frac{g^{\alpha+1}-g^0}{g^{\alpha+1}+g^{\alpha}}\nespb &  0\nespb \\
%0\nespb & 1\nespb & 0\nespb & \frac{g^{\alpha+2}-g^0}{g^{\alpha+2}+g^{\alpha}}\nespb \\
%\frac{1}{g^{2\alpha+1}-g^0}\nespb &  \frac{1}{g^{2\alpha+1}-g^1}\nespb &  \frac{1}{g^{2\alpha+1}-g^2}\nespb & \frac{1}{g^{2\alpha+1}-g^1}\frac{g^{2\alpha+1}-g^0}{g^{2\alpha+1}+g^{\alpha}} \nespb \\
%\frac{1}{g^{2\alpha+2}-g^0}\nespb &  \frac{1}{g^{2\alpha+1}-g^2}\nespb &  \frac{1}{g^{2\alpha+2}-g^2}\nespb & \frac{1}{g^{2\alpha+2}-g^1}\frac{g^{2\alpha+2}-g^0}{g^{2\alpha+2}+g^{\alpha}} \nespb \\
%\frac{1}{g^{2\alpha+3}-g^0}\nespb &  \frac{1}{g^{2\alpha+1}-g^3}\nespb &  \frac{1}{g^{2\alpha+3}-g^2}\nespb & \frac{1}{g^{2\alpha+3}-g^1} \frac{g^{2\alpha+3}-g^0}{g^{2\alpha+3}+g^{\alpha}}
%\end{pmatrix}
\end{align*}
\normalsize

By construction, any $k$ rows of $\Phi$ are invertible; and for a given finite field and $k$, it is easy to check computationally that all $n$ values of $\Lambda$ are different and that all $n$ possible $d\times{}d$ sub-matrices of $\Psi$ are invertible. The construction we give is valid for $k \in \{2...39\}$ in the default GF($2^8$) from GF-Complete~\cite{Plank2013} and $k \in \{2...64\}$  in GF($2^{16}$). Notice that the number of matrices to check computationally is small so that checking all the conditions for all $k$ takes 0.6 seconds in GF($2 ^8$) and 19 seconds for checking up to $k=64$ in GF($2^{16}$). 

As shown in Table~\ref{tab:sparse}, the non-systematic codes are slightly sparser than the non-systematic vanilla codes (\emph{e.g.}, 85~\% instead of 75~\% of zeros for $k=8$). However, the main gain comes for  systematic codes, which are significantly sparser than systematic vanilla codes (\emph{e.g.}, 75~\% instead of 0~\% for $k=8$) thus leading to faster computation. This is important since most codes deployed are systematic.

\begin{table}[t]%
\caption{Sparsity (\emph{i.e.}, percentage of 0) of the generator ($G$ for non-systematic and $G''$ for systematic).} %
\label{tab:sparse} %
\footnotesize %
\centering %
\begin{tabular}{|l||c|c|c|c|} %
\hline
Code (n=2k-1, d=2k-2) & k=4  & k=8 & k=16  \\ \hline\hline
Vanilla PM MSR (non-systematic) & 50 & 75 & 87 \\\hline
Sparse PM MSR (non-systematic) & 64 & 85 & 93 \\\hline
\hline
Vanilla PM MSR (systematic) & 0 & 0 & 0\\\hline
Sparse PM MSR (systematic) & 50 & 75 & 88 \\\hline
\end{tabular} %
\end{table} %

An important advantage of the sparser codes concerns repair. The matrices used for repair (lines  of $\Phi$) contains some zeros (instead of none for vanilla codes). Hence, it implies that helper nodes only have to read the part of their data to be multiplied by non-zero coefficients (\emph{i.e.}, 1 block for a failure of the $\alpha$ first nodes, and $\alpha$ blocks otherwise) instead of all their data (\emph{i.e.}, $\alpha$ blocks). Hence, the disk bandwidth of non-failed devices is preserved. The reduction is on average $\frac{k-2}{2k-2}$. For $k=8$ (resp. $k=4$), the disk bandwidth is reduced by 43\% (resp. 33\%). For large values of $k$, it approaches 50\%.

\section{Evaluation}
\label{sec:eval}
In order to evaluate the computational cost of product-matrix regenerating codes, we implemented them using GF-Complete~\cite{Plank2013} (for the specific version) and Jerasure 2.0~\cite{Plank2014} (for the linearized version). We run our benchmark on a Xeon E5-2640 (2.5 Ghz), which supports SSE (up to 4.2) and AVX instruction sets. Hence, as PM codes require small finite fields (\emph{e.g.,} $GF(2^8)$ is sufficient for $k \le 32$), the libraries can leverage the split-table technique~\cite{Li2013, Anvin2011} relying on SIMD for efficient multiplications. All benchmarks correspond to single threaded computations (\emph{i.e.}, use a single core) and are averaged over 1,000 runs.

For each operation, we separated the total running time into \emph{(i)} initialization time (\emph{e.g.}, building generator or encoding matrix, inverting it) and \emph{(ii)} time needed to apply the operation to the data (reported as the corresponding throughput\footnote{The encoding/decoding throughput is the time divided by the amount of data to encode/decode (\emph{i.e.}, $k\alpha$). The repair throughput is the time divided by the amount of data to repair (\emph{i.e.}, $\alpha$).} in MB/s). This distinction is important, as in a practical deployment, depending on the operation, the initialization phase could be precomputed and its result stored once for all (\emph{e.g.}, encoding), or must be computed at for each operation as it depends on the failure pattern (\emph{e.g.}, decoding) but can be reused across stripes for large objects.

We evaluate the following codes: %
\begin{itemize}[noitemsep,nolistsep]
\item Vanilla Specific Product-Matrix  (PM Spec. Van.)  are the codes described in the seminal paper~\cite{Rashmi2011}.
\item Sparse Specific Product-Matrix  (PM Spec. Spa.)  are the sparse codes described in Section~\ref{ssec:sparse} using specific algorithms of the seminal paper~\cite{Rashmi2011}.
\item Vanilla Linear Product-Matrix  (PM Lin. Van.)  are the linearized (see Section~\ref{ssec:linear}) version of codes described in the seminal paper~\cite{Rashmi2011}.
\item Sparse Linear Product-Matrix  (PM Lin. Spa.)  are the linearized version (see Section~\ref{ssec:linear})  of sparse codes described in Section~\ref{ssec:sparse}.
\item Reed-Solomon codes (RS Vandermonde)  are Vandermonde-based Reed-Solomon codes implemented in Jerasure and are used as a baseline for the performance evaluation.
\end{itemize}

%Explain methodology. All operations have a fixed cost (matrix inversion, computation of the generator matrix, ...), and cost that depend on the amount of data being processed. The two are averaged over 100 runs. The first cost is showed in seconds (the lower the better), while the second cost is reported in the form of the encoding/repair/decoding throughput (the higher the better).

% 2 pages
%Influence of $k$, influence of block size.

%MBR and MSR.

\begin{figure*}
\subfloat[Throughput  -- non-systematic]{ %
\centering %
\includegraphics[height=0.17\linewidth, trim=0 0 1 0, clip=true]{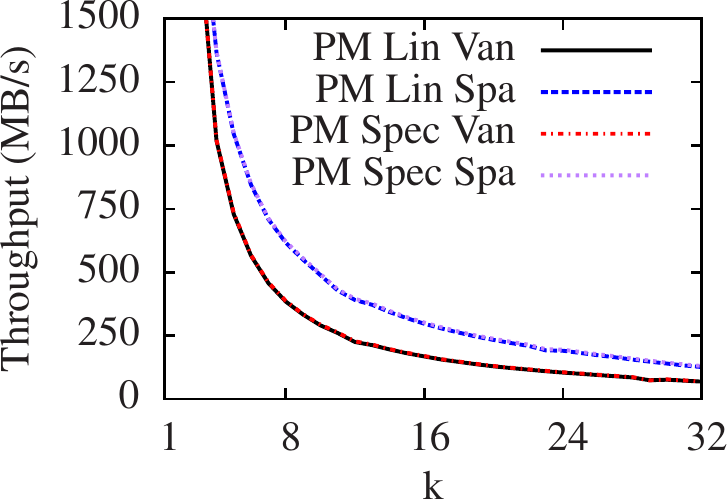} %
\label{fig:enc_bw} %
} %
\subfloat[Throughput -- systematic]{ %
\centering %
\includegraphics[height=0.17\linewidth, trim=17 0 0 0, clip=true]{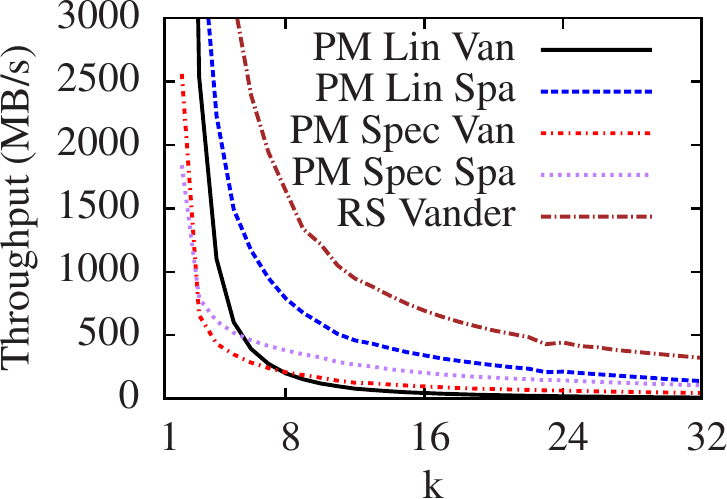} %
\label{fig:encsys_bw} %
} %
\hfill
\subfloat[Initialization -- non-systematic]{ %
\centering %
\includegraphics[height=0.17\linewidth, trim=0 0 1 0, clip=true]{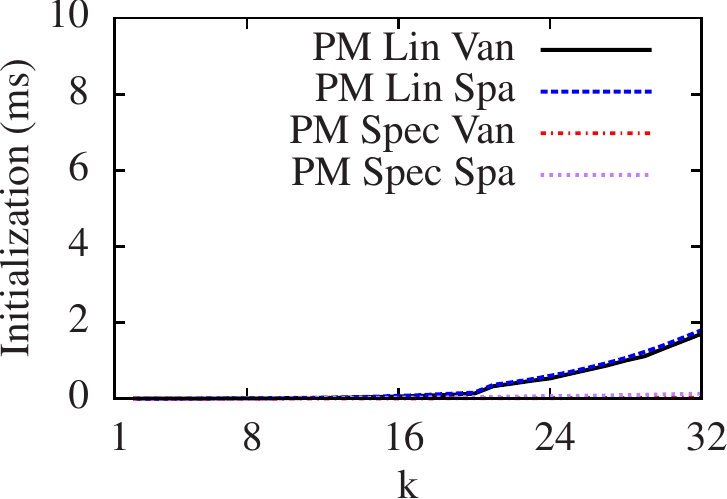} %
\label{fig:enc_pre} %
 } %
\subfloat[Initialization -- systematic]{ %
\centering %
\includegraphics[height=0.17\linewidth, trim=18 0 0 0, clip=true]{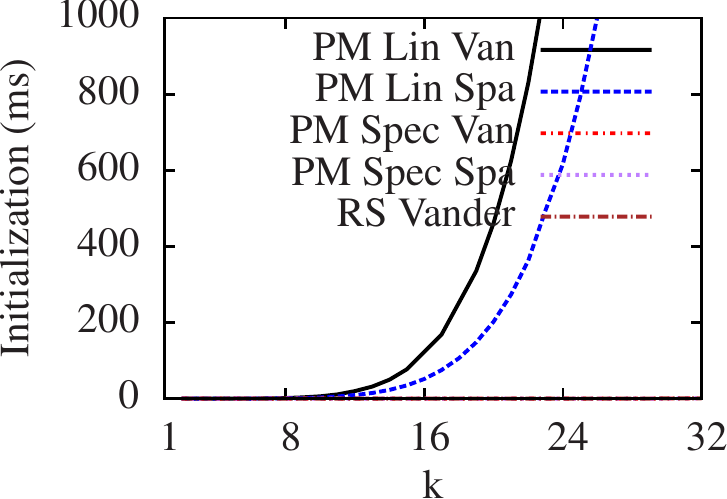} %
\label{fig:encsys_pre} %
} %
\caption{Encoding (regular or systematic) MSR Product-Matrix codes and Reed-Solomon codes. 
%The encoding is split into an initialization step (which can be precomputed once for all), and a step processing the data.
}
\label{fig:enc}
\end{figure*}

\begin{figure*}
\subfloat[Throughput  -- non-systematic]{ % 
\centering %
\includegraphics[height=0.17\linewidth, trim=0 0 1 0, clip=true]{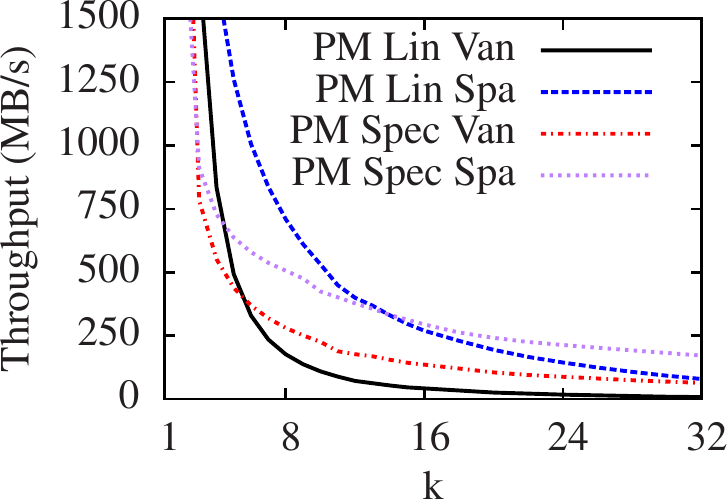} %
\label{fig:dec_bw} %
} %
\subfloat[Throughput -- systematic]{ %
\centering %
\includegraphics[height=0.17\linewidth, trim=17 0 0 0, clip=true]{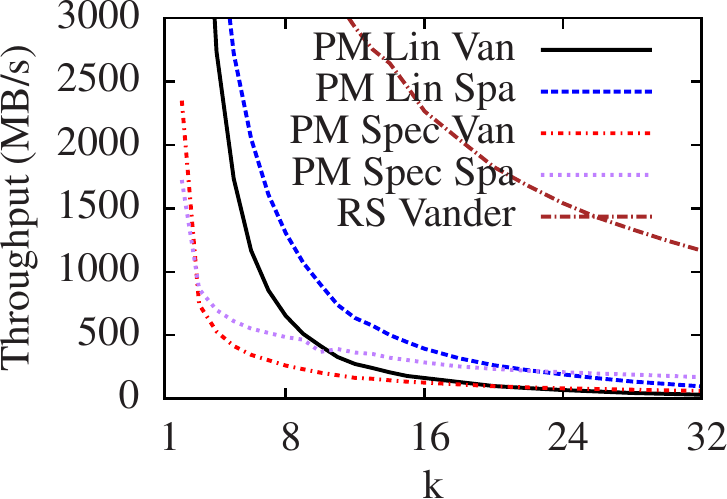} %
\label{fig:decsys_bw} %
} %
\hfill
\subfloat[Initialization -- non-systematic]{ %
\centering %
\includegraphics[height=0.17\linewidth, trim=0 0 1 0, clip=true]{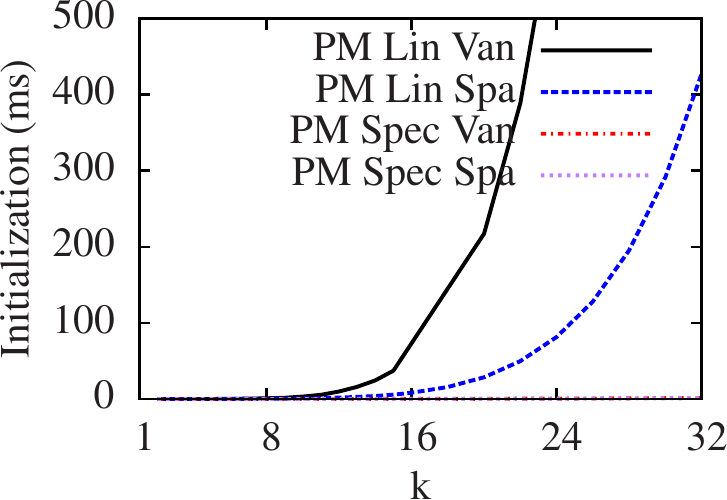} %
\label{fig:dec_pre} %
 } %
\subfloat[Initialization -- systematic]{ %
\centering %
\includegraphics[height=0.17\linewidth, trim=16 0 0 0, clip=true]{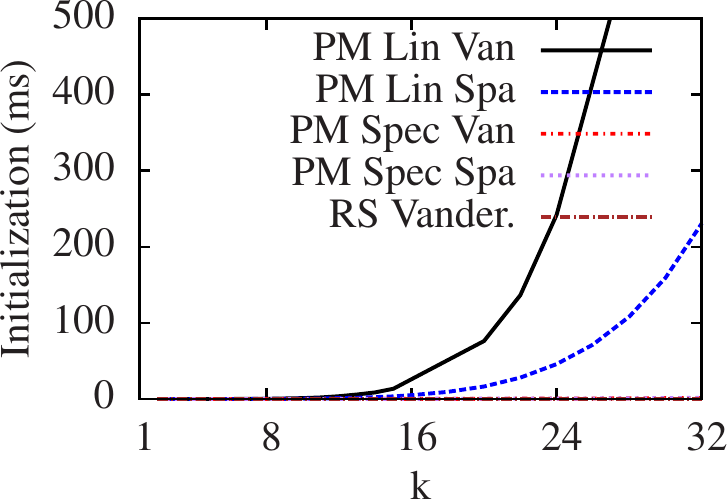} %
\label{fig:decsys_pre} %
} %
\caption{Decoding (regular or systematic) MSR Product-Matrix codes and Reed-Solomon codes.
%The decoding is split into an initialization step independent of the amount of data, and a step processing the data.
}
\label{fig:dec}
\end{figure*}

\begin{figure}[!t]
\subfloat[Non-systematic codes]{ %
\centering %
\includegraphics[height=0.35\linewidth]{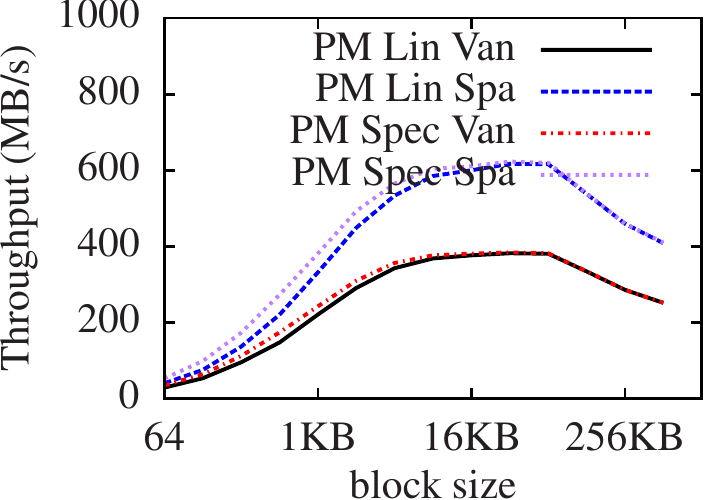} %
\label{fig:enc_nonsys_block} %
}\hfill%
\subfloat[Systematic codes]{ %
\centering %
\includegraphics[height=0.35\linewidth, trim=17 0 0 0, clip=true]{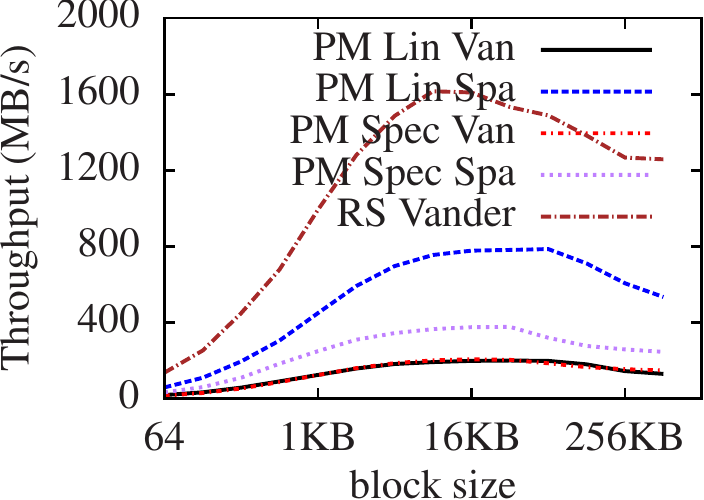} %
\label{fig:enc_sys_block} %
} %
\caption{Impact of block size on MSR encoding.}
\label{fig:block}
\end{figure}

\begin{figure}
\subfloat[Throughput]{ %
\centering %
\includegraphics[height=0.35\linewidth]{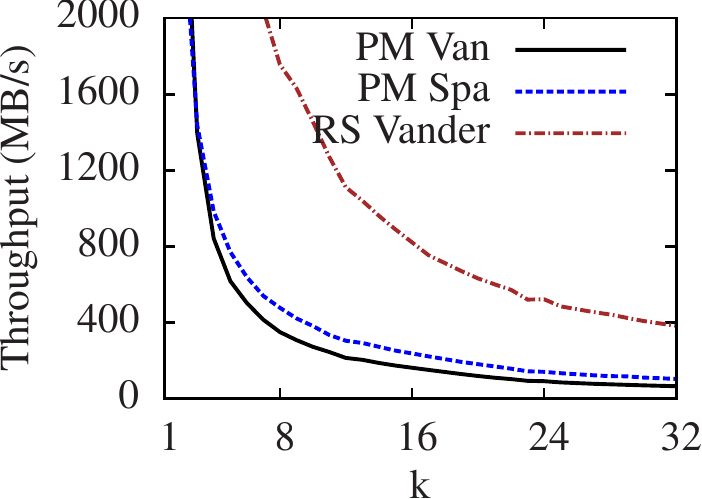} %
\label{fig:rep_thput} %
} %
\vspace{-0.2cm}%
\subfloat[Initialization]{ %
\centering %
\includegraphics[height=0.35\linewidth]{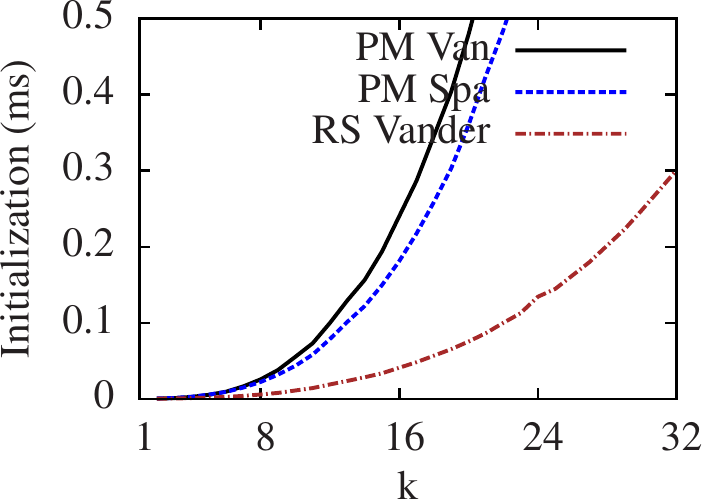} %
\label{fig:rep_init} %
} %
\caption{Repair of MSR PM codes and RS codes.}
\label{fig:rep}
\end{figure}

%\begin{figure}
%\subfloat[Specific or Linearized (Vanilla)]{ %
%\centering %
%\includegraphics[height=0.31\linewidth]{plot/encode_bw_van} %
%\label{fig:enc_lin_van} %
%} %
%\subfloat[Specific or Linearized (Sparse)]{ %
%\centering %
%\includegraphics[height=0.31\linewidth, trim=30 0 0 0, clip=true]{plot/encode_bw_spa} %
%\label{fig:enc_lin_spa} %
%} %
%\caption{Evaluation of the impact of optimizations for encoding systematic product-matrix codes.}
%\label{fig:enc_opt}
%\end{figure}

\subsection{MSR codes}
Figure~\ref{fig:block} shows that a block size of 16 KB gives the best performance for encoding. This is coherent with the size of the L1 cache of the processor, which is 32 KB. Similar results can be observed for decoding, repair and MBR codes. Hence, all the subsequent experiments use 16 KB as the block size.

Figure~\ref{fig:enc} shows the performance of encoding operations.  For $k=8$, a systematic, sparse and linearized PM MSR code allows encoding at 790 MB/s, while Reed-Solomon codes achieve 1640 MB/s and the systematic vanilla PM codes achieve 210 MB/s. Hence, optimized product-matrix codes significantly improve performance over vanilla product-matrix codes; They are a reasonable alternative to Reed-Solomon for practical deployment. 
%It is worth noting that the performance of sparse systematic codes is higher than sparse non-systematic codes because the $k\alpha$ first symbols need not be computed for systematic codes as they are equal to uncoded symbols.
%\footnote{Converting to a bit-matrix and computing a schedule is less efficient than using SIMD for Galois field multiplication, leading to slower systematic encoding at 260 MB/s for vanilla PM codes, and 310 MB/s for sparse PM codes for $k=8$ and blocks of 16 KB.}.  

To achieve these speeds, linearizing the code is useful for systematic codes, as it allows encoding directly without performing the \emph{pre-}decoding step thus doubling the throughput (see PM Lin. Spa. vs PM Spec Spa. on Fig.~\ref{fig:encsys_bw}). However, the linearized version is less efficient than the specific version for large values of $k$ when using the vanilla code (see PM Lin. Van. vs PM Spec Van. on Fig.~\ref{fig:encsys_bw}). This is due to the fact that the generator matrix is not sparse for the systematic vanilla code. Hence, the sparsified version presented in this paper is an important optimization allowing to better leverage the gains from linearization. 
Figure~\ref{fig:encsys_pre} shows that the gain observed in throughput for using linearized systematic product-matrix codes comes at the price of a higher initialization cost (\emph{i.e.}, time needed to compute the systematic generator matrix). Yet, since the generator matrix is constant, it can be computed once and reused for all other encodings. 

Figure~\ref{fig:rep} shows the performance of the repair operation for a single failure. In order to measure the performance, the computation which consist of several independent sub-computations, is performed sequentially on a single core. Sparse PM codes are slightly faster than vanilla codes and are operating at 480 MB/s when Reed-Solomon codes operate at 1830 MB/s.

Figure~\ref{fig:dec} shows the performance of decoding (averaged over many failure patterns with one 1 to $n-k$ failures) using data from systematic devices if available. The initialization cost does not include the generation of the generator matrix, which correspond to the initialization of the encoding operation and can be precomputed once for all. The sparse PM are faster than the vanilla PM. Using the linearized version of sparse PM codes is more interesting for low values of $k$ ($k < 12$ for non-systematic codes, $k<21$ for systematic codes), and for large data (due to the initialization cost). Notice that decoding method is independent of the encoding method: it is possible to encode using the linear algorithm and to decode using the specific algorithm of the seminal paper~\cite{Rashmi2011}.

\subsection{ MBR codes}
In this sub-section, we evaluate the computational performance of PM MBR codes~\cite{Rashmi2011} which are the most versatile minimum bandwidth regenerating codes known. MBR codes have a higher storage overhead than the corresponding MSR codes but lower the network bandwidth during repair. Hence, they are of interest in context where network is a more scarce resource than storage.

\begin{figure}
\subfloat[Throughput]{ %
\centering %
\includegraphics[height=0.35\linewidth]{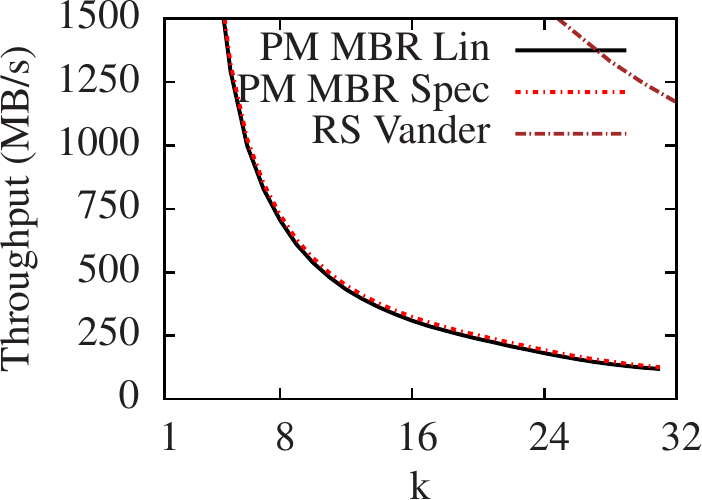} %
\label{fig:enc_mbr_thput} %
} %
\vspace{-0.2cm}%
\subfloat[Initialization]{ %
\centering %
\includegraphics[height=0.35\linewidth]{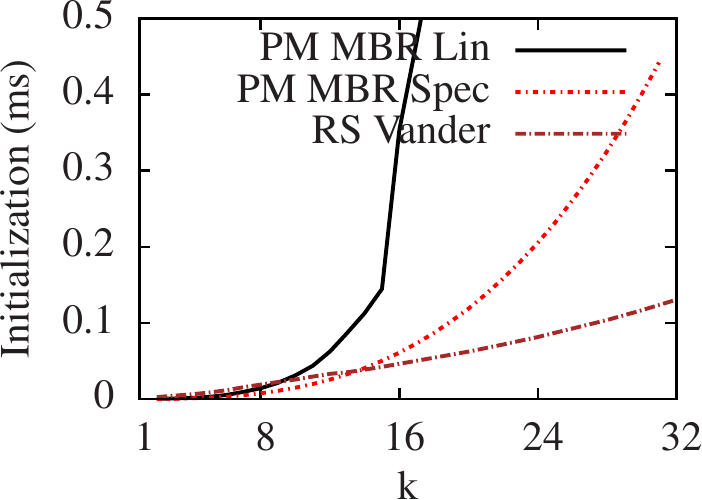} %
\label{fig:enc_mbr_init} %
} %
\caption{Encoding of MBR PM codes and RS codes.}
\label{fig:enc_mbr}
\end{figure}

\begin{figure}
\subfloat[Throughput]{ %
\centering %
\includegraphics[height=0.35\linewidth]{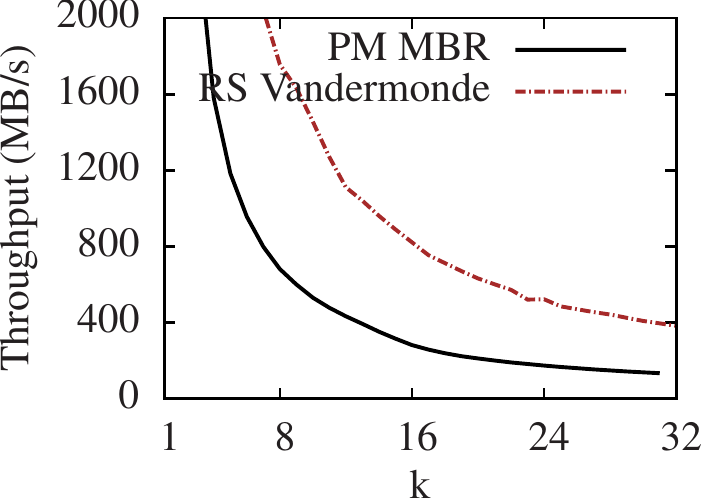} %
\label{fig:rep_mbr_thput} %
} %
\vspace{-0.2cm}%
\subfloat[Initialization]{ %
\centering %
\includegraphics[height=0.35\linewidth]{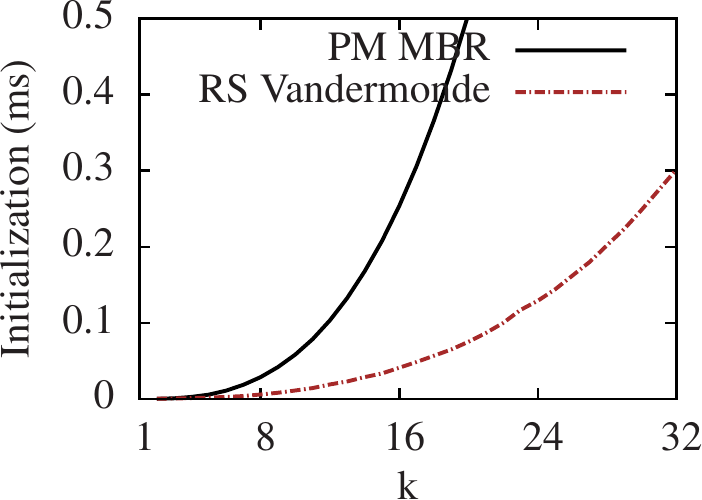} %
\label{fig:rep_mbr_init} %
} %
\caption{Repair of MBR PM codes and RS codes.}
\label{fig:rep_mbr}
\end{figure}

\begin{figure}[!t]
\subfloat[Throughput]{ %
\centering %
\includegraphics[height=0.35\linewidth]{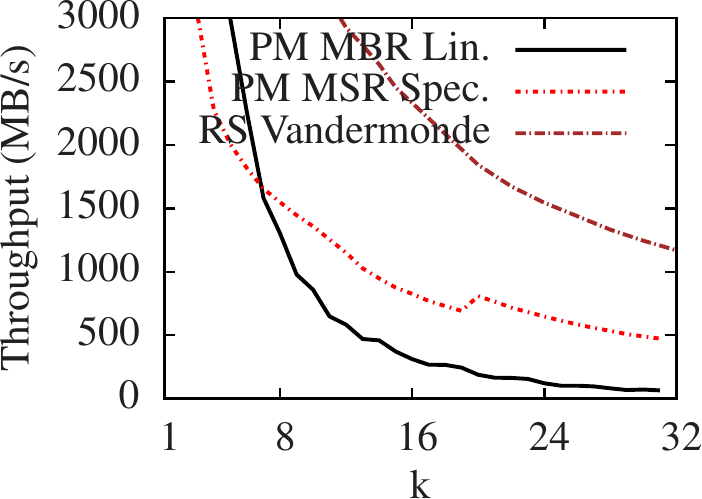} %
\label{fig:dec_mbr_thput} %
} %
\vspace{-0.2cm}%
\subfloat[Initialization]{ %
\centering %
\includegraphics[height=0.35\linewidth]{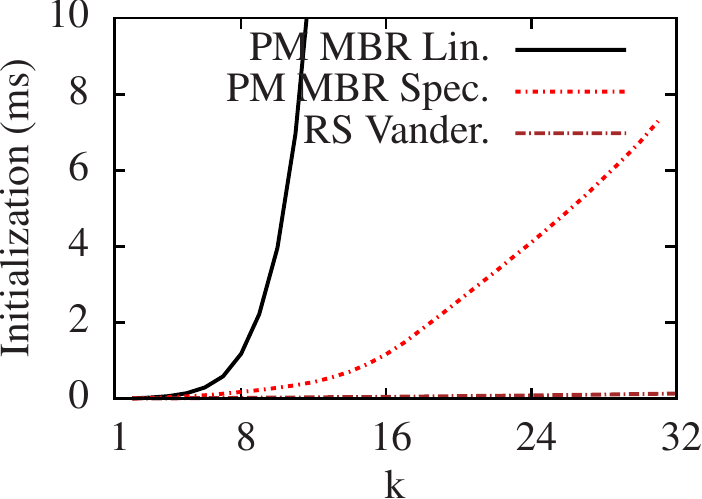} %
\label{fig:dec_mbr_init} %
} %
\caption{Decoding of MBR PM codes and RS codes.}
\label{fig:dec_mbr}
\end{figure}

Figure~\ref{fig:enc_mbr} shows the encoding performance of the PM MBR codes, which are naturally sparse and systematic. The PM MBR codes achieve 725 MB/s for $k=8$, thus having a limited overhead when compared to Reed-Solomon codes (1640 MB/s). This speed is comparable to systematic sparse PM MSR codes.

Figure~\ref{fig:rep_mbr} shows the repair performance of the PM MBR codes. The PM MBR codes achieve 681 MB/s for k=8 which is slightly faster than the PM MSR codes and remain reasonable when compared to the 1830 MB/s of Reed-Solomon codes.

Figure~\ref{fig:dec_mbr} shows the decoding performance of the PM MBR codes. In this case, the linearized version is faster only for $k<7$. Indeed, the specific algorithm for decoding factorizes a lot of computations thus being more efficient. The slight advantage of the linearized version for low values of $k$ comes from the fact that it is possible to selectively decode only a part of the data if only few systematic devices have failed. When $k$ becomes larger, the failure patterns considered (between $1$ and $n-k$ failures) have a stronger impact on the systematic devices making this optimization less efficient. Decoding PM MBR codes is faster than all PM MSR: Indeed, the code structure is simpler and the  redundancy helps the decoding.

% % Graph:
% %   - prepare time / process time
% %   - encode / repair / decode
% %   - k = 8, bs= 8 bytes ... 128 kbyes
% %   - k = 3...32 bs = 16 ko
% %   - PM, Linear PM,  RS
% %
% %  3 x 2   = 6 plots  //  3 curves   for MBR
\vspace{-1eX}

\section{Conclusion}
We presented new sparse product-matrix regenerating codes that encode systematically at 790 MB/s for typical settings (\emph{i.e.} $k=8$), four times faster than vanilla product-matrix regenerating codes~\cite{Rashmi2011} (210 MB/s). Thanks to this significant improvement, regenerating codes achieve half the throughput of Reed-Solomon codes (1640 MB/s). The achieved performance is the result of a good interaction between the linearization of the systematic code that allows to collapse the pre-processing and the encoding; and the sparse structure of the new product-matrix regenerating codes we define. Additionally, sparse codes lower the impact of repair on disks for non-failed devices (43\% less reads for $k=8$).

Beside this new code, we also report numerous throughputs for existing systematic and non-systematic MSR codes and MBR codes, for decoding and repairing. Throughputs reported give insight on the CPU penalty of using regenerating codes and show that it remains limited when compared to Reed-Solomon codes. Throughput reported are an order of magnitude higher than previous studies, thus highlighting that regenerating codes can be used in practical system with little CPU penalty.

%The sparsified product-matrix codes we propose clearly outperform the vanilla codes for all operations. It should be noted that linearization is needed to get the most of them for systematic encoding. Regarding decoding, our measures show that the choice of the procedure (linearized) depends on the size of the file to be decoded because of the initialization time and on $k$. Nevertheless, decoding is not an essential operation when considering systematic codes as read can be performed directly, or using the repair procedure (degraded read) when there is a single failure.

\vfil
\pagebreak
{ %\footnotesize
\bibliographystyle{acm}
\bibliography{hs}
}

 \end{document}